# Using In-Memory Encrypted Databases on the Cloud

Francesco Pagano
Department of Information Technology
Università degli Studi di Milano
Milano, Italy
francesco.pagano@unimi.it

Davide Pagano
School of Engineering
Politecnico di Milano
Milano, Italy
davide1.pagano@mail.polimi.it

*Abstract*—Storing data in the cloud poses a number of privacy issues. A way to handle them is supporting data replication and distribution on the cloud via a local, centrally synchronized storage. In this paper we propose to use an in-memory RDBMS with row-level data encryption for granting and revoking access rights to distributed data. This type of solution is rarely adopted in conventional RDBMSs because it requires several complex steps. In this paper we focus on implementation and benchmarking of a test system, which shows that our simple yet effective solution overcomes most of the problems.

I. INTRODUCTION

Storing sensitive data in the cloud may lead to security fault when it resides on untrusted servers. To solve this issue, a distributed approach was presented in [1], where agents share confidential data in a secure manner using simple grant-and-revoke permissions on shared data. The additional step was the implementation of a distributed DBMS with row-level encryption capabilities to enable a strong access control to records, allowing revocation of rights. This solution is not frequent in literature because of its inherent slowness. In this paper we present a real implementation of such software and describe how we solved the performance problems.

We first describe a schematic model that we introduced in a previous paper (section II), then, after taking a survey on cryptography in databases (section III), granularity in database-level encryption (section IV) and in-memory databases (section V), we describe our solution (section VI), that we implemented and benchmarked (section VII).

II. THE DISTRIBUTED ARCHITECTURE

*A. The model*

Hereon, we will use the term *dossier* to indicate a set of correlated information. Our data model is informally represented in Fig. 1. To simplify the discussion, we introduce the following assumptions:

- Each dossier has only one owner;
- Only the dossier's owner can change it.

These assumptions permit the use of an elementary cascade synchronization in which the owner will submit the changes to the receivers.

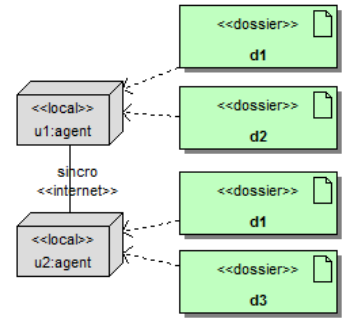

Figure 1. The model

In the model, each node represents a local, single-user application/database dedicated to an individual user ($u_n$). The node stores only the dossiers that $u_n$ owns. Shared dossiers (in this example, $d_1$) are replicated on each node. When a node modifies a shared dossier, it must synchronize with the other nodes that hold a copy of it.

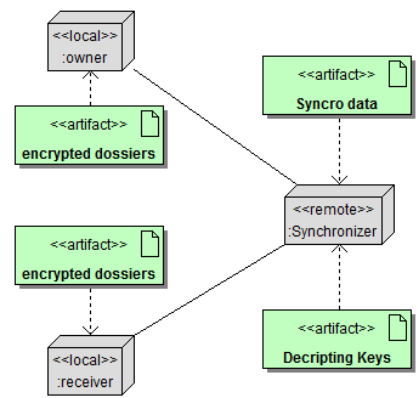

Figure 2. The distributed architecture

Our solution (Fig. 2) consists of two parts: a trusted client agent and a remote untrusted synchronizer on the cloud.

The client maintains local data storage where:

- The dossiers whom she owns are (or at least can be) stored as plaintext;
- The others, instead, are encrypted, each using a different key.

The Synchronizer stores the keys to decrypt the shared dossiers owned by the local client and the modified dossiers to synchronize. No information is in clear-form: dossiers are encrypted using the keys, which, in turn, are encrypted using the receivers' public keys. When another client needs to decrypt a dossier, she must connect to the Synchronizer and obtain the corresponding decryption key. The data and the keys are stored in two separate entities and therefore none can access information without the collaboration of the other part.

## III. CRYPTOGRAPHY IN DATABASES

Confidentiality, integrity and availability are the main properties of database protection. Confidentiality has been defined by the International Organization for Standardization (ISO) in ISO-17799[1] as "ensuring that information is accessible only to those authorized to have access"; data integrity assures that none can modify the information without a trace; availability provides access to data by authorized users in a reasonable time. Along the years, a lot of ACP (Access Control Policy) have been defined, based on database model (*relational* rather than *object*) and policy control (i.e., DAC-Discretionary Access Control, RBACC-Role-Based Access Control, MAC-Mandatory Access Control). Traditionally, ACPs are based on the assumption that the DBA (DataBase Administrator) is trusted, but it is not assured in the outsourced data centers and in the cloud, where the platform-as-a-service (PaaS) provider is external to data owner. A solution to this problem is that the DBMS treats only raw-data, encrypted in such a way that DBA (or another intruder) cannot read the information. There are three main categories of database encryption [4]:

- Storage level encryption
- Database level encryption
- Application level encryption

### A. Storage-level encryption (SLE)

Data is encrypted either at the file level (NAS/DAS) or at the block level (SAN) [5]. A short while ago, Toshiba has released a hardware implementation of SLE, a family of hard drives - called Self-Encrypting Disk. The system is based on the Opal specifications of Trusted Computing Group, supports native encryption AES 256 and can automatically delete its contents if not used by the rightful owner. This encryption is not selective; it encrypts an entire support or portions of support. It prevents theft of storage but it is unsuitable for preventing unauthorized access by a honest-but-curious system administrator. On the other hand, it is entirely transparent to the system, so it needs no database modification.

### B. Database-level encryption (DLE)

DLE secures data as it is written to and read from a database. The encryption is applied to the db at various granularities, such as database, tables, columns (most frequently), and rows. It can be related with some logical conditions for selecting affected data, too. The cons are:

- DLE is not transparent to application as SLE, so it involves some modifications to the indexed encrypted data and in stored procedures and triggers;
- The system is slowed down by the encryption overhead;
- Usually, it is not a defense from the curious DBAs.

### C. Application-level encryption (ALE)

In this case, data is encrypted/decrypted by the application that generates it. Plain-text data is made available only at client side, while data sent over the network is encrypted. This scheme usually involves returning larger result sets to the client, which are then filtered at client side, when decrypted. To accomplish this result, applications need to be modified and the network traffic increases.

## IV. GRANULARITY IN DATABASE-LEVEL ENCRYPTION

The most common solution for data protection is DLE, which can have different types of granularity:

- database
- tables
- columns
- rows

### A. Database

In this case, the whole database is encrypted using only one key, as if it was a single file. The cons of this encryption are:

- It doesn't allow to define different privileges on each table;
- The schema definition becomes particularly complex;
- The system performance suffers considerable degradation (an improvement can be achieved with appropriate caching);
- Its effectiveness is closely linked to the degree of confidence with which the master key is kept.

For these reasons, the database granularity solution is seldom used.

### B. Tables

A specific key encrypts each table separately. Performances are better than the previous solution, but still very far from those of a clear-text database, because encrypting an existing table can be very slow. The definition (and enforcement) of integrity constraints, foreign keys and indexes are very complex.

### C. Columns

All the data in a column (or set of columns) of a table is encrypted with the same key. This is the solution adopted by most DBMS suppliers, as it allows encrypting only sensitive data. However, it needs to build ad-hoc indexes customized for

---

[1] ISO/IEC 17799, Jan 4, 2009

31

the expected queries (again, at the expense of performance). With this approach, it is also not possible to define access privileges on "horizontal" portions of a table such as row sets (e.g., allowing access only to rows with id> 100), as it is awkward to encrypt rows with different keys depending on the user. This type of mechanism usually relies on third-party applications, or otherwise it is delegated to instruments such as triggers or stored procedures.

*D. Rows*

Each single row in a table is encrypted using a different key. The main advantage of this technique is the capability to define access control on a subset of data (rows) of a table basing on the distribution of decryption keys. Let's assume that we have a table that includes the data of all students in a university and we want to grant access to the secretary's office of each course only to data of students enrolled in that course. If we are using database or table-level encryption, we would have to create a view for each course and grant the rights to the corresponding secretary's office, with the problems outlined above (also, data stays readable by the DBAs). Using column-level encryption, the permissions must be specified at the field level and, unless appropriate indexes or cumbersome procedures are implemented (which may also expose the data to inference or statistical attacks), it would be impossible to make the information instantly accessible to authorized users. Using row-level encryption, instead, it is possible to make available to the authorized user the keys (or the key) that can be used to decrypt only the allowed rows. This technique, beside to ensure a better management of access permissions, prevents any kind of statistical analysis on the table. In a normal RDBMS, however, this technique has significant disadvantages in terms of performance and functionality: querying would be possible only through the construction of appropriate indexes for each column of the table (with a considerable waste of resources both in terms of time and space), while the constraints and foreign keys would be almost unusable. Another major issue concerns the management of keys: row-level encryption could potentially lead to the generation and maintenance (and / or distribution) of a key for each row of each table encrypted with this method. To solve (or reduce) the problem, we can use some techniques of key management, such as:

- Broadcast (or Group) encryption[13]: rows are divided into equivalence classes, based on recipients. Every class is encrypted using an asymmetric algorithm where the encryption key is made in a way that each recipient can decrypt the information using only its own private key. Either the public and the private keys are generated by a trusted entity.

- Identity Based Encryption [11]: it bounds the encryption key to the identity of recipient. Each recipient generates by itself a key pair used to encrypt/decrypt information.

- Attribute Based Encryption [12]: it bounds the encryption key to an attribute (a group) of recipient. Each recipient receives by a trusted entity the private key used to decrypt, while the encryption key is calculated by the sender.

However these techniques are complex and therefore conventional RDBMSs don't use encryption at the row-level.

V. IN MEMORY DATABASES

"An in-memory database (IMDB also known as main memory database system or MMDB and as real-time database or RTDB) is a database management system that primarily relies on main memory for computer data storage."[2] It is interesting noting that, while a conventional database system stores data on disk but caches it into memory for access, in an IMDB the data resides permanently in the main physical memory and there is a backup copy on disk [14]. "In-memory databases have recently become an intriguing topic for the database industry. With the mainstream availability of 64-bit servers with many gigabytes of memory a completely RAM based database solution is a tempting prospect to a much wider audience."[3] IMDBs are intended either for personal use (because they are comparatively small w.r.t. traditional databases), or for performance-critical systems (for their very low response time and very high throughput). They use main memory structures, so they need no translation from disk to memory form, and no caching and they perform better than traditional DBMSs with Solid State Disks. Usually, the use of volatile memory-based IMDBs supports the three ACID properties of atomicity, consistency and isolation, but lacks support for the durability property. To add this when non-volatile random access memory (NVRAM) is not available, IMDBs use a combination of transaction logging and primary database check-pointing to the system's hard disk: they log changes from committed transactions to physical medium and, periodically, update a disk image of the database. Having to write updates to disk, the write operations are heavier than read-only. Logging policies vary from product to product: some leave the choice of when to write the application on file, others do all the checkpoints at regular intervals of time or after a certain amount of data entered / edited.

TABLE I. IMDBs PROS AND CONS

| Pros | Cons |
|---|---|
| Fast transactions | Complexity of durability |
| No translation | |
| High reliability | |
| Multi-User concurrency | |

Obviously, the limitation of this type of database is related to the amount of RAM on computer hosting the db. But given their nature, IMDBs are well suited to be distributed and replicated across multiple nodes to increase capacity and performance. The proposed solution works around this limitation: not having a single central database containing the whole data, we preferred to give one database for each client application. This database contains only owned data, while external data will be added (or removed) via the Synchronizer, based on access permissions. To minimize cryptography overhead, we encrypt only rows "received" by other nodes, while rows owned by the local node are stored in clear form.

---

[2] http://en.wikipedia.org/wiki/In-memory_database
[3] http://www.remote-dba.net/t_in_memory_cohesion_ssd.htm

Well-known open solutions of IMDB are Apache Derby, HyperSQL (HSQLDB) and SQLite. For our implementation, we chose to use the open source HyperSQL rel. 2.0.

*A. HyperSql*

HyperSQL[4] is a pure Java RDBMS. Its strength is, besides the lightness (about 1.3Mb for version 2.0), the capability to run either as a Server instance either as a module internal to an application (in-process). A database started "in-process" has the advantage of speed, but it is dedicated only to the containing application (no other application can query the database). For our purposes, we chose server mode. In this way, the database engine runs inside a JVM and will start one or more "in-process" databases, listening requests from processes in the local machine or remote computers. For interactions between clients and database server, we can use three different protocols:

- HSQL Server: the fastest and most used. It implements a proprietary communication protocol;
- HTTP Server: it is used when access to the server is limited only to HTTP. It consists of a web server that allows JDBC clients to connect over http;
- HTTP Servlet: as the Http Server, but it is used when accessing the database is managed by a servlet container or by an application servlet (e.g. Tomcat). It is limited to using a single database.

There are different types of databases (called catalogs) that can be created with HyperSQL, that differ in the methodology adopted for data storage:

- res: this type of catalog provides for the storage of data into small JAR or ZIP files;
- mem: data is stored completely in the machine's RAM, so there is no persistence of information outside of the application life cycle in the JVM;
- file: data is stored in files residing into the file system of the machine.

In our work we used the last type of catalog. A catalog file can use up to six files on the file system for its operations. The name of these files consists of the name of the database plus a dot suffix. Assuming we have a database called "db_test", the files will be:

- db_test.properties containing the basic settings of the DB;
- db_test.log: used to periodically save data from the database, to prevent data loss in case of a crash;
- db_test.script: containing the table definitions and other components of the DB, plus data of not-cached tables;
- db_test.data: containing the actual data of cached tables. It can be not present in some catalogs;
- db_test.backup: containing the compressed backup of last ".data" file, that may be not present in some catalogs;
- db_test.lobs: used for storing BLOB or CLOB fields.

Besides these files, HyperSQL can connect to CSV files.

A client application can connect to HyperSQL server using the JDBC driver (.Net and ODBC drivers are "in late stages of development"), specifying the type of database to access (file, mem or res). HyperSQL implements the SQL standard either for temporary tables either for persistent ones. Temporary tables (TEMP) are not stored on the file system and their life cycle is limited to the duration of the connection (i.e. of the Connection object). The visibility of data in a TEMP table is limited to the context of connection used to populate it. With regard to the persistent tables, instead, HyperSQL provides three different types of tables, according to the method used to store the data:

- MEMORY: it is the default option when a table is created without specifying the type. *Memory table* data is kept entirely in memory, while any change to its structure or contents is recorded in .log and .script files. These two files are read at the opening of database to load data into memory. All changes are saved when closing the database. These processes can take a long time in the case of tables larger than 10 MB.
- CACHED: when this type of table is chosen, only part of the data (and related indexes) is stored in memory, thus allowing the use of large tables at the expense of performance.
- TEXT: the data is stored in formatted files such as .csv.

In our work, we use MEMORY tables. The Loader and the Serializer are the main parts of HyperSQL that we analyzed and modified. They are the mechanisms that load the data from text files at the opening and save them to the database at closing.

*1) Loader*

We suppose that the client connects to the DBMS using instructions like:

```
Class.forName("org.hsqldb.jdbcDriver" );
Connection c = DriverManager.getConnection(
  "jdbc:hsqldb:file:myDB", "SA", "");
```

Having used a catalog of *file* type, the static method newSession() of class org.hsql.DatabaseManager is called. Its task is to open the database or to connect to it (if it is already opened). org.hsql.Database is the class that represents the instance of the database in memory, so this is the root of all data structures designed to contain the information of the database. Once the database is loaded into memory, two fundamental classes are used for the parsing of text files: org.hsqldb.ParserCommand (for management of sessions and statements) and org.hsqldb.Scanner (for the recognition of individual SQL tokens). The class responsible for maintaining the database (related to the session) is org.hsqldb.SessionData, whose main attributes are:

```
private final Database database;
private final Session session;
PersistentStoreCollectionSession persistentStoreCollection;
```

*PersistentStore* is the data structure that contains all rows in a database table. Specifically, this is an interface implemented by using different classes depending on the type of table

---
[4] www.hsqldb.org

33

represented: in our case we use MEMORY tables, so that the affected class is the org.hsqldb.persist.RowStoreAVLMemory. When the *Database* object is created, particularly at the invocation of method reopen(), the class org.hsqldb.persist.Logger, which is the class that represents the interface for I/O to and from text files of the database, is instantiated. The starter method of Logger class is openPersistence(), which will open the specified database (if the database is new, the related text files are created). The class org.persist.Log is instantiated after verifying the integrity of the *.properties* file. Our focus is on method open() of this class which checks the status of the Database (if it was closed properly, if it was modified, and so on) and then instantiates the class org.hsqldb.scriptio.ScriptReaderText to read the *.script* file using the method *readAll(Session s)*. The class org.hsqldb.rowio.RowInputTextLog is used to read a single line of the database and the object that represents a row in the database is the object Row. Two methods of class *ScriptReaderText* are invoked:

- readDDL(): reads the DDL statements and initialize a class RowInputTextLog for each line read from the *.script* file.
- readExistingData(): it extrapolates the values of each single line, initializes the row and adds it to the *PersistentStore*;

Because of the database file structure, we need to look for *Insert* statements to find the rows of a table. When one of these statements is encountered, it is managed by the method *processStatement(Session s)* of *ScriptReaderText* class. For each field in the row, it checks whether it is primary key and determines the data type, then the value of the field is read by the method *readData (DataType t)* of *RowInputTextLog* class.

*2) Serializer*

The serializer is the module responsible for saving the modified data into .script and .log files. Changes are initially written in .log file and moved to the .script file, when a *shutdown* command is issued. Each database table is represented by an instance of class *org.hsqldb.Table*, comprising: data structures for the management of content, methods for creating a new table, and operations of insert/select rows. When inserting a new row, the method *insertSingleRow()* of the *Table* class is invoked; the first step is to create a new *Row* object for caching data in memory, which is done by the method *getNewCachedObject (Session s, Object [ ] data)* of *PersistentStore* class. Memory-type tables are kept in a balanced tree structure (AVL) implemented in the class *org.hsqldb.persist.RowStoreAVLMemory*. Once a node (i.e. the row being inserted) is built and added to the AVL (this operation involves several checks on the contents of the fields and of integrity constraints), HyperSQL writes the row into the buffer and then transfers it to the text file (data is written to the .log file until shutdown of the database). To perform this task, the Logger class utilizes the method *writeInsertStatement(Session s, Table t, Object [] data)*, and the method *writeInsertStatement()* of the *Log* class. Writing to the file is done using the class org.hsqldb.scriptio.ScriptWriterBase (more precisely, in case of memory-type tables, the *ScriptWriterText* subclass). The method *writeRow(Session s, Table t, Object [] data)* of *ScriptWriterText* class writes data to a text buffer and, at the end of the procedure, transfers it to the file. The buffer (which is only a byte[ ]) is encapsulated in the class *RowOutputBase* (more precisely, in case of memory-type tables, the *RowOutputTextLog* subclass), which extends the *HsqlByteArrayOutputStream* and provides methods to transform any type of data for serializing it into the buffer. Once writing to the buffer is completed, the method *writeRowOutToFile()* of *ScriptWriterText* class is used, which calls the method *write(byte [ ] b)* of the class *OutputStream* to write into the output stream of .log file. When shutting down the database, method *writeScript*() of *Log* class is invoked with the following tasks: creating temporary file for writing .script file, loading each element of the database into memory and writing it to the file by executing the *flush()* of the *OutputStream* connected to the file.

VI. IMPLEMENTED SOLUTION

*A. Client side*

On the client side, using IMDBs, we have only two interactions between each local agent and the Synchronizer.

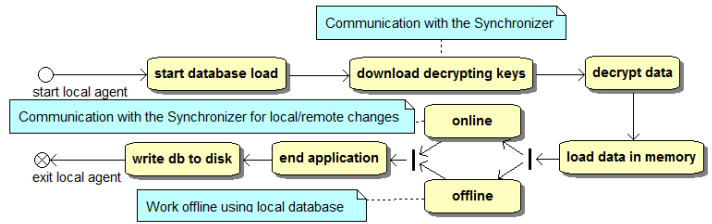

Figure 3. Client's state diagram

Note that the client, after the first communication with the Synchronizer, can run offline. We modified the classes included in file hsqldb.jar to handle the encryption. The basic idea was to manage encryption in the .log and .script text files. The rows that are owned by the local client are stored in clear-text, while the shared rows "granted" by other owners are stored encrypted. The values contained in tables are stored in form of SQL insert:

INSERT INTO table_name(field_1, field_2, …, field_n)
VALUES(value_1, value_2, …, value_n)

Earlier, to obtain control access granularity at the field level, we encrypted field by field. This way, the text contained in the database file was in the form of:

INSERT INTO table_name(field_1, field_2, …, field_n)
VALUES(pk, encrypted_value_2, …, encrypted_value_n)

The primary key pk must be in clear-text, since it is used to retrieve the decrypting keys from the central Synchronizer. We dropped this idea because it requires changing the I/O code for each possible database type and an attacker may obtain some information such as table, primary key and number of rows. The current solution is to encrypt the whole row by AES symmetric algorithm. The encryption overhead is lower than the previous solution and all information is hidden to curious eyes. To relate the encrypted row (stored locally) to the decrypting key (stored in the remote Synchronizer), we use a new key (id_pending_row). The encrypted row is prefixed by a clear-text header containing the id_pending_row delimited by "$" and "@". The encrypted value is then stored in a hexadecimal representation, so a generic row is of the form:

```
$27@5DAAAED5DA06A8014BFF305A93C957D
```

*1) Load time*

At load time, the .script file will contain clear-text and encrypted rows, e.g.:

```
INSERT INTO students(id,name) VALUES(12,'Alice');
INSERT INTO students(id,name) VALUES(31,'Bob');
$27@5F3C25EE5738DAAAED5DA06A80F305A93C95A
$45@5DA67ADA06AAED580FA914BF3C953057D387F
INSERT INTO students(id,name) VALUES(23,'Carol');
```

The class whose task is reading the file and loading the appropriate data in memory is *ScriptReaderText*.

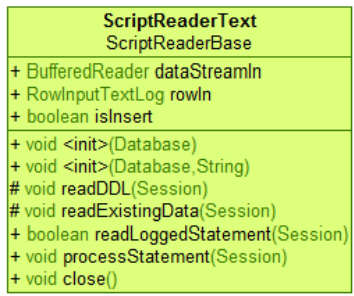

Figure 4. ScriptReaderText class' UML

The *readLoggedStatement* method parses each line of text in the .log or .script files and forwards the result to the *processStatement* method, which loads data into memory. We changed the *readLoggedStatement* method to make a preprocessing: if it finds a record header (enclosed between $ and @) in the text line, it extracts the *id_pending_row_received*. Using this id, the client requests to the central Synchronizer the related decoding key, which it uses to decrypt the entire text line and to proceed with normal HyperSQL management. If the decoding key is unavailable, the text line is temporarily discarded (it is not deleted if it was not received for communication problem with the Synchronizer).

*2) Save time*

The class ScriptWriterText manages the write operations in .log and .script files. The affected methods are *writeRow* and *writeRowOutToFile*. The former deals with building the string that will be written into the text file (INSERT INTO ....) which corresponds to the in-memory data. A *Table* instance contains the information about the table structure (table name, field names, types of data, constraints, etc.). The values of fields are in an array of *Object*. The SQL *insert* is written in a text buffer that is stored in the .script file by the method *writeRowOutToFile*. Because each table has an *id_pending_row_received* column, we modified the *writeRow* method to check if the row is owned or shared by another user. In the latter case (*id_pending_row_received* not null), the custom *writeRowOutToFileCrypto* method is used instead of the original *writeRowOutToFile* method. *WriteRowOutToFileCrypto* uses the parameter *id_pending_row_received* to query the related symmetric encryption key from the Synchronizer, needed to encrypt the whole buffer. The result is a hexadecimal sequence which is prefixed by the below header with the id_pending_row_received.

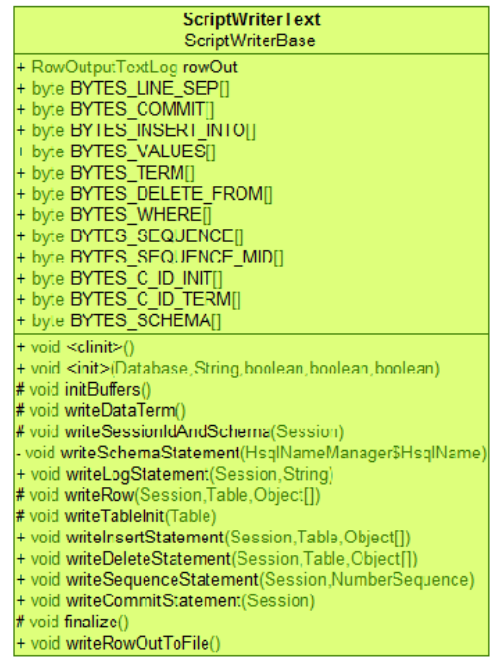

Figure 5. ScriptWriterText class' UML

*3) Changes*

We can alter the original HyperSQL in three ways:

- subclass original classes and override the affected methods
- change the original code directly
- use code injection by aspect programming

The first is the cleanest method, but it implies collaborating with the support team of HyperSQL to implement some interfaces containing new methods to add new features in subclass. Lacking it, we had to revert to the second way. The third, which is less invasive and more maintainable, forces to have AspectJ compiler (or equivalent) in the client library. The changes regarded the classes *ScriptReaderText* and *ScriptWriterText* only (without changing the classes that use them). These are the classes that deal with I/O to and from the .script file. They are very mature and stable, so we think that a simple substitution of its .class files is sufficient to alter hsqldb.jar.

*B. Server side*

When a data owner adds or updates a row in the local database, it needs to propagate it to all the related users utilizing a central Synchronizer acting as a mailbox at server side, in the cloud. It uses a simple database with the following tables:

- Users: containing, among others, the id and public key of each user;

- Pending Rows: containing the rows that are added/modified in the local owner's database, until they are delivered to destination. A unique row_id is automatically assigned to each pending row. Additional information includes: submission date, sender and receiver. The changed row is stored in encrypted form in the field encrypted_row;
- Decrypting keys: contains the keys that are used to decrypt the pending rows. Additional information includes: sender, receiver, expiry date, id_row.

At change time, the owner (client side) must:

- serialize the row;
- generate a symmetric key to encrypt it;
- encrypt the row;
- encrypt the key using the public keys of receivers;
- send the encrypted row and the decoding keys to the receivers.

Because we store the serialized row, we need not worry about columns data types. The Synchronizer uses RMI to expose its services to clients. The services are grouped in three interfaces:

- KeyInterface with methods related to encryption keys: depositKey, deleteDecriptingKey, getDecriptingKeyByIdPendingRow, getPublicKeyByUser;
- SynInterface with methods for sharing the rows: sendRow, getPendingRowForUser, getAllUsers, resendRow;
- RegistrationInterface to register and manage users: registerUser, SelectUserById, selectUserByIdAndPassword.

## VII. PERFORMANCES

In contrast to the usual row-level encryption, which needs encryption/decryption at every data access, our solution uses these heavy operations only when communicating with Synchronizer, with a clear advantage, especially in the case of rarely modified databases.

### A. Read operations

The system uses decryption only at start time, when records are loaded from the disk into the main memory. Each row is decrypted none (if it is owned by local node) or just once (if it is owned by a remote node), so this is optimal for read operations. Each decryption implies an access to the remote Synchronizer to download the related decrypting key and, eventually, the modified row.

### B. Write operations

Write operations occur when a record is inserted / updated into the db, with no overload until the client, when online, explicitly synchronizes data with the central server. At this moment, for each modified record, the client needs to:

- generate a new (symmetric) key
- encrypt the record
- dispatch the encrypted data and the decrypting key to the remote synchronizer

### C. Benchmark

The test application we wrote uses our modified HyperSQL driver and interacts with the other clients through our Synchronizer. It performs these distinct activities:

- Creation of database and sample tables
- Population of tables with sample values
- Sharing of a portion of data with another user
- Receipt of shared dossiers from other users
- Opening of the newly created (and populated) database

The application receives three parameters:

- Number of dossiers
- Number of clients involved in sharing
- Percentage of shared dossiers

To minimize communication delay, the central Synchronizer and the clients ran on the same computer. For testing purpose, it was sufficient to use only two clients (to enable data sharing). The application was compared with an equivalent one having the following differences:

- It uses the unmodified HyperSQL driver
- It doesn't share data with other clients
- When populating the database, it creates the same number of dossiers than the previous application; after benchmarking, however, it adds the number of shared dossiers, resulting in the same final number of dossiers.

We benchmarked the system using single-table dossiers of about 200 bytes, in two batteries of tests; the first with 20%, and the second with 40% of shared dossiers, which numbered from 1,000 to 500,000. The results are represented by the graphs in Fig. 6-8. It is worth noting that the overhead percentage of the modified solution rapidly decreases (with 100,000 dossiers it is around 10%), either in the first battery of tests (Fig. 6), and either in the second (Fig. 7). In the tests, the total delay (load + create + populate + receive) is linear in the number of dossiers and is limited, even with a huge number of dossiers (Fig. 8). Local results can be slightly altered by external events not preventable (e.g., garbage collector).

### D. Results

The delay of the system is tightly bound to communications effort with central Synchronizer. Computing overhead is limited to just one encryption per record at write time and no more than one decryption per record at read time. Since we use symmetric encryption, these operations are very fast. The benchmark demonstrates that the delay is substantially concentrated in database opening, while the subsequent use does not involve additional delays, compared to the unmodified version.

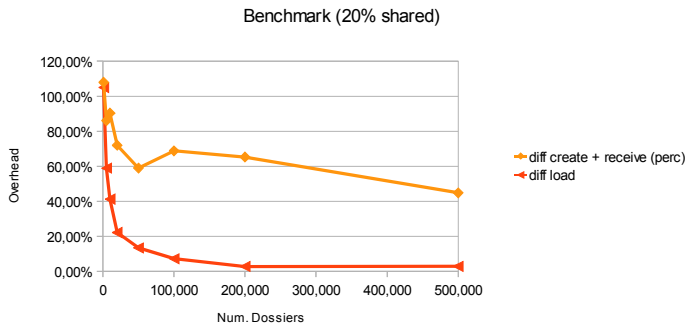

Figure 6. Overhead when 20% of dossiers are shared

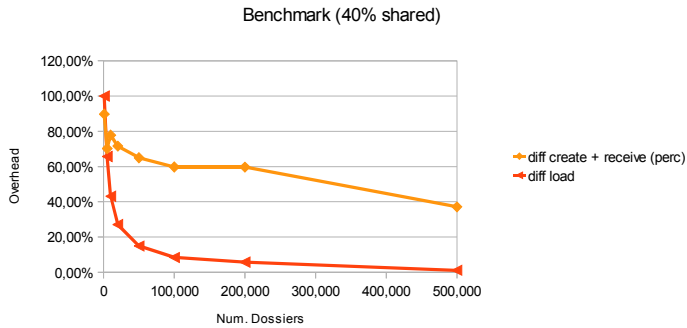

Figure 7. Overhead when 40% of dossiers are shared

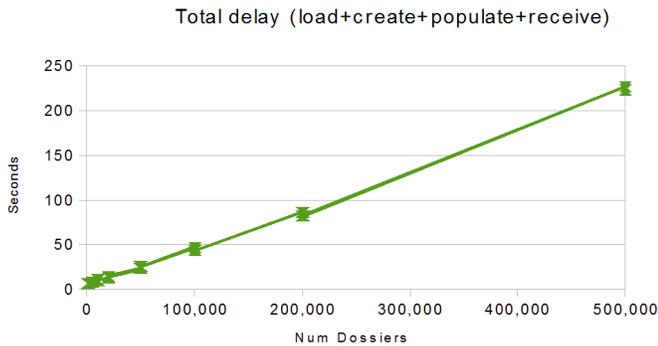

Figure 8. Total delay

## VIII. CONCLUSION

In this paper, using IMDBs, we presented a simple solution to row-level encryption of databases. It can be used in the cloud to manage very granular access rights in a highly distributed database. This allows for stronger confidence in the privacy of shared sensitive data. An interesting field of application is the use in (business) cooperative environments, e.g. professional networks. In these environments, privacy is a priority, but low computing resources don't allow the use of slow and complex algorithms. IMDBs and our smart encryption, instead, achieve the goal in a more effective way.

## IX. FUTURE WORK

We want to test the system in case of large population of users in the cloud. We are working to reduce the number of communications between local nodes and synchronizer using a form of group encryption. We are going to compare the complexity of the naïf solution with the group encryption effort to evaluate which are the parameters that affect the performance of the two alternatives.

## ACKNOWLEDGMENT

We wish to thank Ernesto Damiani, professor at Università di Milano, for helpful comments. Marco Di Paola contributed within his graduation dissertation to the design and implementation of this application.